\documentclass[a4paper,10pt]{article}
\usepackage[utf8]{inputenc}
\usepackage{graphicx}
\usepackage{subcaption}
\usepackage[english]{babel}
\usepackage {natbib}
\usepackage{array}
\usepackage{lscape}
\usepackage{float}
\usepackage{xcolor}
\usepackage{booktabs}
\usepackage[left=3.75cm, right=3.75cm, top=2cm, bottom=2cm]{geometry}
\usepackage[super]{nth}
\usepackage{fourier} 
\usepackage{makecell}
\usepackage{keyval}
\usepackage{amsmath}
\usepackage{fixltx2e}
\usepackage{longtable}
\usepackage{csquotes}
\usepackage{times}
\usepackage{elcvia}
\usepackage[english]{babel}

\input{psfig.sty}

\newcommand{\vol}{0}            
\newcommand{\num}{0}            
\newcommand{\begpag}{\thepage}  
\newcommand{\finpag}{7}         
\newcommand{\yearp}{2020}       

\title{Change in Artificial Land Use over time across European Cities: A rescaled radial perspective 
\footnote[0]{Correspondence to: $<$paul.kilgarriff@liser.lu or p.kilgarriff1@gmail.com$>$ \vspace{0.2cm}}
\footnote[0]{This is a draft chapter. The final version will be available in "Handbook on Entropy, Complexity and Spatial Dynamics: A Rebirth of Theory?"}
\footnote[0]{edited by Aura Reggiani, Laurie A. Schintler and Danny Czamanski, forthcoming 20xx, Edward Elgar Publishing Ltd.}
\footnote[0]{The material cannot be used for any other purpose without further permission of the publisher, and is for private use only.}
}

\author{Paul Kilgarriff$^{*}$, Rémi Lemoy$^{§}$ and Geoffrey Caruso$^{*+}$ \\ \\
\centerline{\small \em $^{*}$ Urban Development \& Mobility, Luxembourg Institute of Socio-Economic Research (LISER), Luxembourg} \\
\centerline{\small \em $^{§}$ University of Rouen, IDEES Laboratory UMR 6266 CNRS, France} \\
\centerline{\small \em $^{+}$ Dept. of Geography and Spatial Planning, University of Luxembourg, Luxembourg} \\ \\
\centerline{\small \today}
       }
      
\begin{document}

\maketitle

\pagestyle{myheadings}

\markboth{\centerline{\small \it Kilgarriff, Lemoy \& Caruso / 
           Change in Artificial Land Use over time across European Cities: A rescaled radial perspective 
           }}
         {\centerline{\small \it Kilgarriff, Lemoy \& Caruso / 
           Change in Artificial Land Use over time across European Cities: A rescaled radial perspective
           }}
\hrulefill

\begin{abstract}

Seen from a satellite, observing land use in the daytime or at night, most cities have circular shapes, organised around a city centre. A radial analysis of artificial land use growth is conducted in order to understand what the recent changes in urbanisation are across Europe and how it relates to city size. We focus on the most fundamental differentiation regarding urban land use: has it been artificialised for human uses (residence or roads for instance) or is it natural, or at least undeveloped? Using spatially detailed data from the EU Copernicus Urban Atlas, profiles of artificial land use (ALU) are calculated and compared between two years, 2006 and 2012. Based on the homothety of urban forms found by \citet{Lemoy2018}, a simple scaling law is used to compare the internal structure of cities after controlling for population size. We firstly show that when using the FUA definition of cities, a kind of Gibrat’s law for land use appears to hold. However, when we examine cities internally, this is no longer clear as there are differences on average between city size categories. We also look at further city groupings using regions and topography to show that artificial land use growth across European cities is not homogeneous. Our findings have important implications relative to the sustainability of cities as this evidence is pointing towards increasing urban sprawl and stagnant growth in urban centres across cities of all sizes. It also has theoretical implications on the nature of sprawl and its scaling with city size.
 
\vspace{0.5cm}
\noindent
{\em Key Words}: Spatial analysis, radial analysis, 
land use, scaling laws, urban sprawl
\end{abstract}
\hrulefill

\section{Introduction}

Cities are changing both in terms of population and land use. While studies exist that examine the relationship between population growth and city size, few studies examine the relationship between artificial land use (ALU) growth and city size and fewer still examine the relationship at the intra-urban level. By examining the relationship between growth in ALU (i.e. urbanisation) and city size, we test whether ALU growth exhibits Gibrat’s Law. We add to previous examinations of Gibrat’s Law for cities by examining the intra-urban structure of change. Cities are analysed at various distances to the central business district (CBD) after those distances are made comparable across cities of different size. Cities are also grouped using context or historical characteristics, captured by the coastal/non-coastal location of the city or the year the country joined the European Union (EU). An examination of the evolution of urban areas is important as this is where the majority of people live \citep{Ioannides2009}. For example the 293 cities used for this study, represent 44\% of the population of the EU-27. It is important to examine the regularity of cities and seek a statistical explanation for the hierarchy of cities \citep{pumain2006hierarchy}. Then it is important to examine whether urbanisation patterns are statistically similar for large cities and small cities since these patterns impact urban sustainability, and a policy might typically need to favour smaller or larger cities. 

    ``\emph{Even if Gibrat’s model remains generic and universal, it can by no means be accepted as a 'purely' stochastic process requiring no further explanation. On the contrary, it should be enriched by reference to historical context and trends}''
\\[5pt]
\rightline{{\rm --- \citet{pumain2006alternative}}}

Gibrat’s 1931 book "Les Inégalités Économiques" \citep{Gibrat1931} used several variables (income, population, wealth) to examine why despite the growth process being considered stochastic, the variables often exhibit skewness or kurtosis, despite the central limit theorem \citep{akhundjanov2019gibrat} where a normal distribution would be expected. Gibrat proposed \citep{Gibrat1931} that the proportional rate of growth is independent of size and gives rise to a distribution that is log-normal. More simply a city with population 10 million is as likely to double in size during a given period as a city with population 500,000 \citep{Mansfield1962}. The consequences of Gibrat and Zipf illustrate that the drivers of agglomeration economies are the same for all cities \citep{Batty2008a}.

Several studies have examined Gibrat for a range of different cities and countries, with mixed results but showing city size and population growth are independent of each other and hence follow Gibrat’s law at least to a certain extent \citep{robson1973urban,pumain1982dynamique,deVries1984european,moriconi1993urbanisation,guerin1992deux,Eeckhout2004}. \citet{robson1973urban} examined UK urban agglomerations >2,500 inhabitants between 1801-1901 using ten time intervals and found that Gibrat held for the most part with a slight deviation, as the level of variance in growth rates among smaller cities was higher compared to large cities. \citet{pumain1982dynamique} examined French towns >2,500 inhabitants over five time intervals between 1831-1975. Gibrat’s law could only be verified during periods of slow growth. When annual growth exceeded 2\%, an increasing trend of growth rate and city size was found. A temporal autocorrelation of growth rates is also found during high growth periods \citep{pumain1982some}. 

Over a longer time period (100 years) Gibrat’s law semi-holds but is weaker in the short term. Over the long term Gibrat’s law is found to be proportional in means but not in variance \citep{Gonzalez-Val2013}. A null hypothesis of the city size distribution being log-normal is only rejected for early decades in the \nth{20} Century \citep{Gonzalez-Val2013}. Although six years is a relatively short period of time some cities have experienced growth $>$10\% over this period. \citet{Eeckhout2004} used a period of ten years to show that cities grow independent of city size and hence follow Gibrat’s law.  \citep{akhundjanov2019gibrat} re-examined the 24 data sets used in Gibrat’s 1931 book using modern computing and found that 18 of the 24 data sets follow a mixture of Pareto and log-normal distributions, with an exponent at the upper tail of less than 2. The data city on city size is found to be indifferent between a power law or log-normal distribution.

\citet{pumain2006alternative} discusses a commonly found deviation from Gibrat: a low and often positive correlation between urban growth rates and city size. Equivalently, if cities are classified according to increasing size there will be an increase in the mean value of growth rates. Two explanations are offered to this deviation, first innovations are often adopted in large cities first with large cities then benefiting from initial advantage. The second explanation relates to the increase in speed of transportation which increases a large cities sphere of influence, stealing market share from smaller surrounding towns and cities.

When testing whether Gibrat’s law holds for cities, studies typically consider each city as a whole. This approach fails to recognise the variation in growth within cities, especially the potential effect of the physical pattern of urbanisation (ALU) or of varying population density levels \citep{guerois2008built,schneider2008compact,jiao2015urban}. Differences in growth rates between the core, suburbs and periphery are typically ignored. City definition and how they are defined is therefore important when examining cities \citep{Rozenfeld2011}. Cities can be defined using a functional approach \citep{OECD2013}, as a network such as the City Clustering Algorithm \citep{Rozenfeld2008} or using administrative units such as a city boundary. When defining cities using arbitrary city limits or urban boundaries, issues such as the Modified Aerial Unit Problem (MAUP)  \citep{Openshaw1984} and interactions between closely located cities \citep{thomas2018city} are encountered.

To understand the urban expansion process a deeper analysis of the internal structure of cities is required. Cities are here defined as a function of distance to the CBD using a radial analysis. Similar to other studies \citep{Walker2018,Wilson2012} the location of the city hall is used as the centre point of cities. This point tends to coincide with the principal residential centre \citep{Griffith2007}. Assuming a city has one dominant centre point at its core assumes cities are monocentric. This assumption of monocentric cities and use of distance based measures is long established in urban geography and economics \citep{Alonso1964,Clark1951,Fujita1989,McDonald1989,VonThunen1875}. Even within polycentric cities or when important e sub centres exist, though, there is a main centre with the highest population density \citep{Griffith2007}). In a study of US metropolitan regions, monocentric was the most prevalent \citep{Arribas-Bel2014}. One of the advantages of using a radial analysis, is the ability to examine the complex two-dimensional intra-urban structure of a city in a one-dimensional space. This approach is compatible with traditional urban economic theory such as the \citet{VonThunen1875} and Alonso-Mills-Muth models \citep{Alonso1964,Mills1972,Muth1969a}, which despite their simplifications have the merit of stressing the importance of central accessibility costs on locational decisions. The rescaling methodology used for this study has been found to be compatible with the Alonso model \citep{Delloye2018} corrected for land use. One of the fundamental determinants of the urban desnity of a city is the distance to the CBD and the trade-off that occurs between price of land and cost of commuting/accessing the city center \citep{Brueckner1987}. Using a distance based approach enables us to understand urban sprawl and how to mitigate it. We aim to discover which cities are sprawling and how sprawl varies depending on city size further grouping types.

In Europe between 1990 and 2006, $\approx$1,000 km$^2$ of land (40\% of the total area of the country of Luxembourg) was converted per year to be used for housing, roads and industry. Soil sealing is the covering of soil with impermeable materials such as concrete \citep{Prokop2011}, a process which is rarely reversed. Soil sealing as a result of artificial land use has several associated negative impacts such as; loss of water retention, loss of biodiversity and unsustainable living patterns \citep{Prokop2011}. It is one of the main factors threatening the state of soil in Europe \citep{Jones2012}. In 2006, 100,000 km$^2$ or 2.3\% of all EU land was sealed, with levels above 5\% in Netherlands, Belgium and Germany. Growing levels of urban expansion is a challenge for cities as they seek to expand sustainably and efficiently. With Europe already highly urbanised any increase in ALU is likely to occur outside of the city core. 

Poor planning can lead to increasing levels of urban sprawl. Between 1990-2015 there were increasing levels of built-up land and decreasing density \citep{denis2020more}. Such urban sprawl makes provision of utilities and mass transit more difficult. The European Environment Agency (EEA) has described urban sprawl as the pattern of low-density expansion of large urban areas into surrounding areas which are mostly agricultural \citep{EEA2006}. Increasing urban sprawl is the uptake of built-up areas, dispersed over a given landscape with low utilisation intensity in the built-up area \citep{Jaeger2014}. Urban sprawl has several negative effects such as loss of agricultural land, increasing fragmentation, destruction of ecosystems, higher transport costs and increases in greenhouse gas emissions \citep{EEA2016}.

The urban expansion process is problematic as it removes land from agriculture, green spaces and nature and is often irreversible. It is the over expansion of ALU relative to the population that is the issue for cities \citep{Brueckner2000}. Over expanding of ALU can also put increasing pressure on existing public services and result in capacity issues. This urban expansion process can take various forms such as sprawling patterns or fragmented patterns. Land conversion from agriculture to artificial (residential or commercial) will occur when society values artificial land more than agricultural land. Cities where agricultural land is highly productive are typically more compact \citep{brueckner1983economics,oueslati2015determinants}. For most of Europe, the value of residential land will surpass the value of agricultural land with exceptions largely due to the topography of land and expensive construction costs. Urban expansion can result from a growing population, rising income and falling commuter costs. Excessive urban expansion may result in several market failures \citep{Brueckner2000}. Failure to account all benefits associated with urban green space. Negative externalities associated with excessive commuting and cities that are too large. Failure to account fully the cost of all public infrastructure associated with urban expansion \citep{Brueckner2000}. Sprawling cities will consume more fuel in transportation but also more land, and infrastructure materials for water, electricity and roads \citep{OMeara1999}.

In 2017 building construction and operations accounted for 36\% of global energy use and 39\% of energy-related carbon dioxide emissions \citep{IEA2018}. In 2013, the world’s urban areas accounted for about 64\% of global primary energy use and produced 70\% of the planet’s carbon dioxide emissions \citep{IEA2016}. If current trends are to continue, combined with the increasing population of cities, by 2050 urban primary energy demand will increase by 70\% accounting for 66\% of global demand, carbon emissions will also increase by 50\% \citep{IEA2016a}. Final energy demand in the buildings and transport sector can be reduced by 60\% through reduced length and frequency of trips and energy saving homes and low carbon fuels. The urban form and density can create the premises for reduced demand for mobility and for greater efficiency of energy use in buildings, including the opportunity to integrate low-carbon district heating and cooling networks with heat generated by low-carbon fuels or waste heat from industrial plants \citep{IEA2016a}.

The change in ALU may vary depending on city size. How fast are small cities growing compared to larger cities? Smaller cities tend to use more land per capita than larger cities. The fast growing, newer cities also tend to use more land per capita compared to the older, slower growing cities \citep{Boyce1963}. This is why an in-depth examination of Gibrat’s law is interesting, not only to examine it at an aggregate level but also at an intra-urban level. Controlling for city size using scaling, enables us to compare the change in ALU for different groups of cities, i.e. small versus large cities. This will be related to issues such as sprawl and urban expansion. The second argument for focusing on Gibrat is where that change happens and whether that ‘where’ is changing with city size. We can examine where the biggest change in ALU is occurring in relation to the CBD. Examining the internal structure of cities enables us to open and look inside the black box of city land use. 

There is a long established literature which uses scaling laws to compare cities \citep{Batty2013,Bettencourt2007,Louf2014}. 
 This chapter utilises a scaling methodology developed by \citet{Lemoy2018} who found that the radial artificial land use profiles of different cities are quite similar if the distance to the city center is rescaled using the total population to an exponent 1/2. This rescaling enables us to hold population constant and compare cities of different population sizes and hence different areas. This analysis examines artificial land use (ALU) growth/change, as opposed to population growth. Population growth demands a certain level of expansion however when this expands beyond the given population growth it becomes a problem. Some cities may be using too much land compared to their size. We know that surface is related to population \citep{Lemoy2018}, it is unclear whether a change in artificial surface is related to population or not, which is the main reason behind examining Gibrat’s law.

This chapter examines ALU growth in Europe between 2006 and 2012. A radial analysis is used to calculate the level of artificial land use at several distances to the city centre. Using the radial scaling law of \citet{Lemoy2018} cities to control for population, we examine change in a systematic manner for ~300 European cities. The compatibility of Gibrat’s law and ALU is investigated first at an aggregate level (city) using artificial land use (ALU) growth/change and the population using the Larger Urban Zone (LUZ) definition of cities. A disaggregated approach is then used to further examine the internal structure of change within cities.


 Analysing where the change occurs within cities as a function of distance to the CBD will inform us of how this change is occurring. If city expansion is not homogenous across distance, where are the highest levels of urban expansion occurring? Are these distances the same across the range of city sizes? 

\section{Methodology}

Cities are examined both at an aggregate and intra-urban level. A radial analysis is introduced to examine the internal structure of cities. Cities are analysed using concentric rings around a single point (co-ordinates of the historic city hall) to represent the CBD. 
A scaling exponent is then used to control for city size. The data used in this analysis is from the Urban Atlas dataset produced by Copernicus land monitoring service \citep{Copernicus2016}. In this section we first describe the data, how ALU is defined then introduce the scaling law used in the analysis.

\subsection{Urban Atlas}

The data used comes from the EU Copernicus Urban Atlas \citep{Copernicus2016}, which is available at a 5m resolution. The boundary of each city corresponds to its Functional Urban Area (FUA). A subset of cities which appear in both the 2006 and 2012 editions are used which leaves 293 cities all located within the EU-27 \& the UK. These cities range in population from 62,000 to 11m (Paris \& London). In defining the CBD of the FUA, the location of the historic city hall is used as the point to represent the CBD. As every city has a city hall this is a method which enables us to perform the radial analysis in a systemic and consistent way \citep{Walker2018,Wilson2012}.

Between the two years of the Urban Atlas, 2006 and 2012, there are some changes to the boundary of the FUA. To ensure a common area between the two years the 2006 FUA is clipped using the 2012 FUA and vice versa, which leaves us with their intersection. This ensures we are using the largest possible area that features in both the 2006 and 2012 Urban Atlas. This new common area is then used to calculate the population for each city using the 2006 EU GEOSTAT 1km$^2$ population grid \citep{Eurostat2012}. 

Artificial land use is examined to measure the increasing/decreasing levels of soil sealing across European cities. Out of the 20 land use categories for 2006 and 24 land use categories in 2012, 12 categories are combined to calculate a measure of artificial land use (ALU); land use classes 1-12 from table \ref{tbl:uacodes}. When we refer to ALU, it is these 12 categories to which we are referring to.

\begin{table}[ht]
\begin{tabular}{|r|l|}
  \hline
\# & Category  \\ 
  \hline
1 & Continuous urban fabric (S.L. : > 80\%)\\
2 &	Discontinuous dense urban fabric (S.L. : 50\% -  80\%)\\
3 &	Discontinuous medium density urban fabric (S.L. : 30\% - 50\%)\\
4 &	Discontinuous low density urban fabric (S.L. : 10\% - 30\%)\\
5 &	Discontinuous very low density urban fabric (S.L. : < 10\%)\\
6 &	Isolated structures\\
7 &	Industrial, commercial, public, military and private units\\
8 &	Fast transit roads and associated land\\
9 &	Other roads and associated land\\
10 & Railways and associated land\\
11 & Port areas\\
12 & Airports\\
13 & Mineral extraction and dump sites\\
14 & Construction sites\\
15 & Land without current use\\
16 & Green urban areas\\
17 & Sports and leisure facilities\\
18 & Arable land (annual crops)\\
19 & Forests\\
20 & Water\\
  \hline
\end{tabular}
\caption{Urban Atlas land use classes}\label{tbl:uacodes}
\end{table}
  
The twelve categories chosen to represent artificial land use are those where buildings are dominant, urban atlas codes starting with '11' or '12'. These are found to have the least amount of variability between the two years and do not suffer from other issues relating to reclassifying facilities and amenities such as for construction and urban green areas. For example a construction site in period t, can cover a larger area compared to the resulting building footprint in period t+1 with the remaining area in one of the other non-urbanised categories.

It is worth noting there are some limitations with the Urban Atlas. The reclassifying land despite no changes occurring can make inter-temporal analysis more challenging as you want to insure the increase in artificial land use being observed is as a result of activity and not because of reclassification. Reclassification between years occurs in the Urban Atlas. In Munich where the biggest re-categorisation of cemeteries occurred, $\approx$11 hectares were converted from green urban areas to an artificial land use category. However this represents only 0.00001\% of total artificial land for Munich. Despite this, we are satisfied our results are not sensitive to these small re-categorisations. As one of the goals of this research is reproducibility, the number of edits made to the original data should be kept to a minimum. For this reason no changes have been made to the master Urban Atlas data.

\subsection{Geoprocessing}

The data from the Urban Atlas is first rasterised into a 20m resolution grid. Each raster cell is given the land use of the polygon at the centre of cell. With the CBD as the centre point concentric rings with a fixed width $\delta=100\sqrt{2}$ $\approx$ 141m are created until the outer edge of the FUA is reached. It is then possible to examine the share of ALU at various distances to the CBD. In addition to rings ALU shares are calculated using discs. As opposed to rings, discs consider the entire area of the concentric circle with radius r, and not just the area of the difference between a circle with radius $r$ and a circle with radius $r+\delta$.

\subsection{Scaling}

This chapter uses a previously discovered homothetic scaling law to transform ALU \citep{Lemoy2018}. The ALU of a city was found to scale with city size measured by its total population in a homothetic manner. More precisely, the total artificial area of a city is proportional to its total population, and the radius of the city scales with the square root of its total population. This is the standard relationship between the area and the side length of a surface in two dimensions (square or disc for instance). We note that homothetic or isometric scaling uses a fixed factor for all parts of the considered system, in comparison to allometry which uses different rates of growth \citep{thompson1917growth,Huxley1932} for different parts of the system.

This homothetic scaling of artificial land use can be expressed with mathematical relations. \citet{Lemoy2018} found that the radial artificial land use profiles s(r) of different cities are quite similar if the distance r to the city center is rescaled to a distance r$'$ given by:

\begin{equation} \label{eq2}
r'=r \times \sqrt{\cfrac{N_{\text{ London}}}{N}}=r \times k
\end{equation}

where N is the population of the city being analysed and N\textsubscript{London} is the population of the largest city in the dataset, used as a reference.

\begin{equation} \label{eq3}
k=\sqrt{\frac{N_{\text{ London}}}{N}}
\end{equation}

is the rescaling factor. For London and Paris, k$\approx$1. How this scaling works is illustrated in example 1 below.

\noindent\fbox{%
    \parbox{\textwidth}{%
Example 1: Worked example of scaling\\
Rescaling Liverpool (UK) using its population (1,371,238) and the reference city of London (UK), population (11,312,174).
Eq. \ref{eq14} shows how the rescaling factor k is calculated:
\begin{equation} \label{eq14}
k=\sqrt \frac{Pop_{\text{ London}}}{Pop_{\text{ Liverpool}}}=\sqrt \frac{11,312,174}{1,371,238}\simeq 2.872
\end{equation}

k is used as the rescaling factor to rescale distances for Liverpool. Eq. \ref{eq15} illustrates this for Liverpool using a distance of 1.2km.
\begin{equation} \label{eq15}
r'= r\times k=1.2\times 2.872= r'=3.456 \text{ km}	
\end{equation}

Result: After rescaling, 1.2 km Liverpool has a similar land use profile to 3.4 km London. At 10 km Liverpool has a similar profile to 29 km London
    }%
}

The radial analysis of ALU in this chapter uses two different measures: rings and discs. These two measures provide different insights on artificial land use: measures in discs study the share of artificial land within a given distance r from the center, while measures in rings study artificial land at a given distance r (more precisely, between r and r+$\delta$, where $\delta$ is the width of the ring).

The surface V(t,r,i) corresponding to a particular land use class i in a disc of radius r at time (year) t is described in eq. \ref{eq4}:

\begin{equation} \label{eq4}
V(t,r,i)=\pi r^2 v(t,r,i)
\end{equation}

Where t is time (year), r is distance (radius) and v(t,r,i) is the share of the disc corresponding to land use class(es) i.

The surface S(t,r\textsubscript{1},r\textsubscript{2},i) of a ring can be seen as the difference between the surfaces of an outer disc of radius r\textsubscript{2} and an inner disc of radius r\textsubscript{1}:

\begin{equation} \label{eq5}
S(t,r_1,r_2,i)=V(t,r_2,i)-V(t,r_1,i)
\end{equation}

where t is time and i is the considered land use class.

\begin{equation} \label{eq6}
s(t,r_1,r_2,i)=\left( \frac{S(t,r_2,i)}{\pi r_1^2 - \pi r_2^2} \right)
\end{equation}

is the share of this same ring corresponding to land use class i. From these measures we can now derive two temporal measures of ALU change. Eq.\ref{eq7} shows relative change in ALU $\Delta$V(t1,t2,r,i), i.e. how much a disc has changed given its previous level of ALU:

\begin{equation} \label{eq7}
\Delta V(t_1,t_2,r,i)=\left( \frac{V(t_2,r,i)-V(t_1,r,i)}{V(t_1,r,i)} \right)
\end{equation}

where $\Delta$ indicates change, V is the surface of a disc, r is the radius, i is the considered land use class, t is time(year). 

And $C(t_1,t_2,r,i)$ is the conversion rate given by

\begin{equation} \label{eq8}
C(t_1,t_2,r,i)=\left( \frac{v(t_2,r,i)-v(t_1,r,i)}{(1-v(t_1,r,i))} \right)
\end{equation}

Note that we subtract the share of a disc from 1 to calculate the total non -artificialised share of a disc. This conversion metric is more effective at showing change in ALU close to the CBD, where the artificial share of ALU is already high. These measures of change can also be computed for rings instead of discs by replacing V by S and v by s. We replace r with r’ if we are examining rescaled distances. Table \ref{tbl:scdist} compares rescaled distance to the equivalent actual distance for a number of city populations.	

\begin{longtable}{|l|r|r|r|r|}
\hline
& \multicolumn{4}{c|}{Rescaled Distance} \\
 & r$'$=15km & r$'$=20km & r$'$=40km & r$'$=60km \\
\hline
Population & \multicolumn{4}{c|}{Actual Distance} \\
\hline
100,000 & 1 & 2 & 4 & 6 \\
250,000 & 2 & 3 & 6 & 9 \\
500,000 & 3 & 4 & 8 & 13 \\
1,000,00 & 4 & 6 & 12 & 18 \\
2,000,000 & 6 & 8 & 17 & 25 \\
5,000,000 & 10 & 13 & 27 & 40 \\
\hline
\caption{Scaling - Comparison of non-scaled and rescaled distances}\label{tbl:scdist}
\end{longtable}

\begin{figure}[H]
    \includegraphics[width=\linewidth]{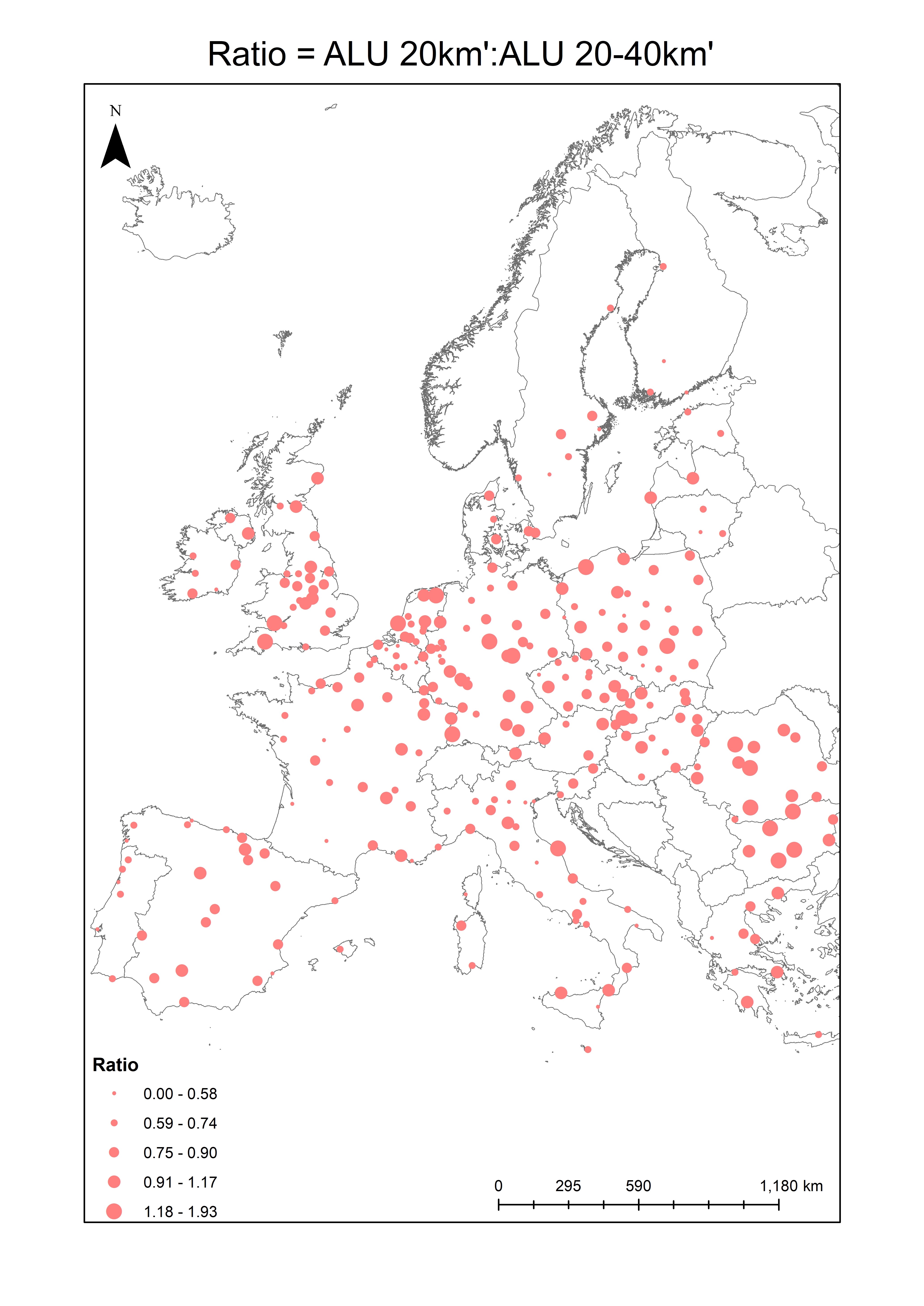}
    \caption{Ratio of ALU r$'$=20km to ALU r$'$=20-40km 2012}
    \label{fig:map}
\end{figure}

In Fig. \ref{fig:map}, we map a ratio of core artificial land to peripheral artificial land computed as:
\begin{equation} \label{eq9}
V(t,r’_1,i)\bigg/S(t,r’_1,r’_2,i) 
\end{equation}
 
 With r$’$\textsubscript{1}=20km and r$’$\textsubscript{2}=40km, that is a disc of radius r$’$\textsubscript{1} and the ring of inner radius r$’$\textsubscript{1} and outer radius r$’$\textsubscript{2}. The higher the ratio the more developed the city is at its core (within r$’$\textsubscript{1}=20km) relative to the periphery (between r$’$\textsubscript{1}=20km and r$’$\textsubscript{2}=40km).
 
\subsection{City categories}

To display the results more effectively, cities are grouped together based on common attributes. Three categories are used; city size, region and share of water. For city size cities are grouped using an adapted version of the OECD size categories \citep{Dijkstra2012}. We adapt the ranges so there are five categories instead of six; small (50k-250k), medium (250k-500k), large (500k-1m), x-large (1m-2m) and xx-large (2m+). As there are only 13 cities with a population below 100k, the small and medium categories from the OECD definitions are amalgamated to form a new small category (50k-250k). Each subsequent category is a doubling of the previous category with the exception of the top and bottom categories. 

From table \ref{tbl:ctypop} we see 44\% of the total EU27 population lived in these 293 cities, with 19\% living in the 24 largest cities with a population over 2 million. Overall 60\% of the sample population live in a city with a population greater than 1,000,000.

\begin{table}[ht]
\begin{tabular}{|l|r|r|r|r|}
  \hline
City Size & Population & No. cities & Share of sample & Share of EU27 total (2011)  \\ 
  \hline
Small & 15,669,921 & 96 & 0.07 & 0.03\\
Medium & 30,830,288 & 84 & 0.14 & 0.06\\
Large & 40,688,516 & 59 & 0.18 & 0.08\\
X-large & 41,093,837 & 29 & 0.18 & 0.08\\
Xx-large & 93,939,287 & 24 & 0.42 & 0.19\\
\hline
Total & 222,221,849 & 293 & 1 & 0.44\\
\hline
\end{tabular}
\caption{Breakdown of cities by city size (2006 population)}\label{tbl:ctypop}
\end{table}

To group cities based on their location and economic status, the year in which the country joined the European Union is used to create three categories; those that founded the European Economic Community (EEC) in 1957 (France, Germany, Italy, Luxembourg, Belgium, Netherlands) (labelled EU-1), countries who joined in the intervening years up to 1996 (UK, Ireland, Spain, Portugal, Greece, Austria, Sweden, Finland, Denmark) (labelled EU-2) and finally new EU member states who joined between 2004-2013 (Poland, Romania, Czech Republic plus another seven) (labelled EU-3). Countries in each of the three groups (EU-1, EU-2, EU-3) share common attributes, the founders category contains European wealthiest countries with many of these countries being heavily urbanised, those who joined in the following years were also advanced economies but located around the edge of Europe such as the UK, Scandinavia and Iberia. The new member states are mostly former communist states and can benefit from the large share of recent EU structural funding.

Differences in typology such as water or elevation may explain some of the differences in ALU and how a city expands and develops influencing its urban form \citep{kasanko2006european}. The final categorisation takes this into account by measuring the share of coastal water surrounding the city, using Corine Land Cover (CLC) data for 2006. Following a radial analysis, the share of water within discs at various distances to the CBD is calculated. r$'$=40km is used along with a cut-off threshold of 10\% share of water, a value used as it is high enough to select only cities with a large body of water such as a sea, ocean or lake. Using a rescaled distance helps to capture only those cities where the share of water limits urban development given their size. 

\section{Results}
The results are divided into two sections. The first section examines the relationship between ALU and city size using different city definitions. The second section analyses the internal structure of cities with the addition of city categories.

\subsection{Gibrat’s Law for land use}

\begin{figure}[H]
    \includegraphics[width=\linewidth]{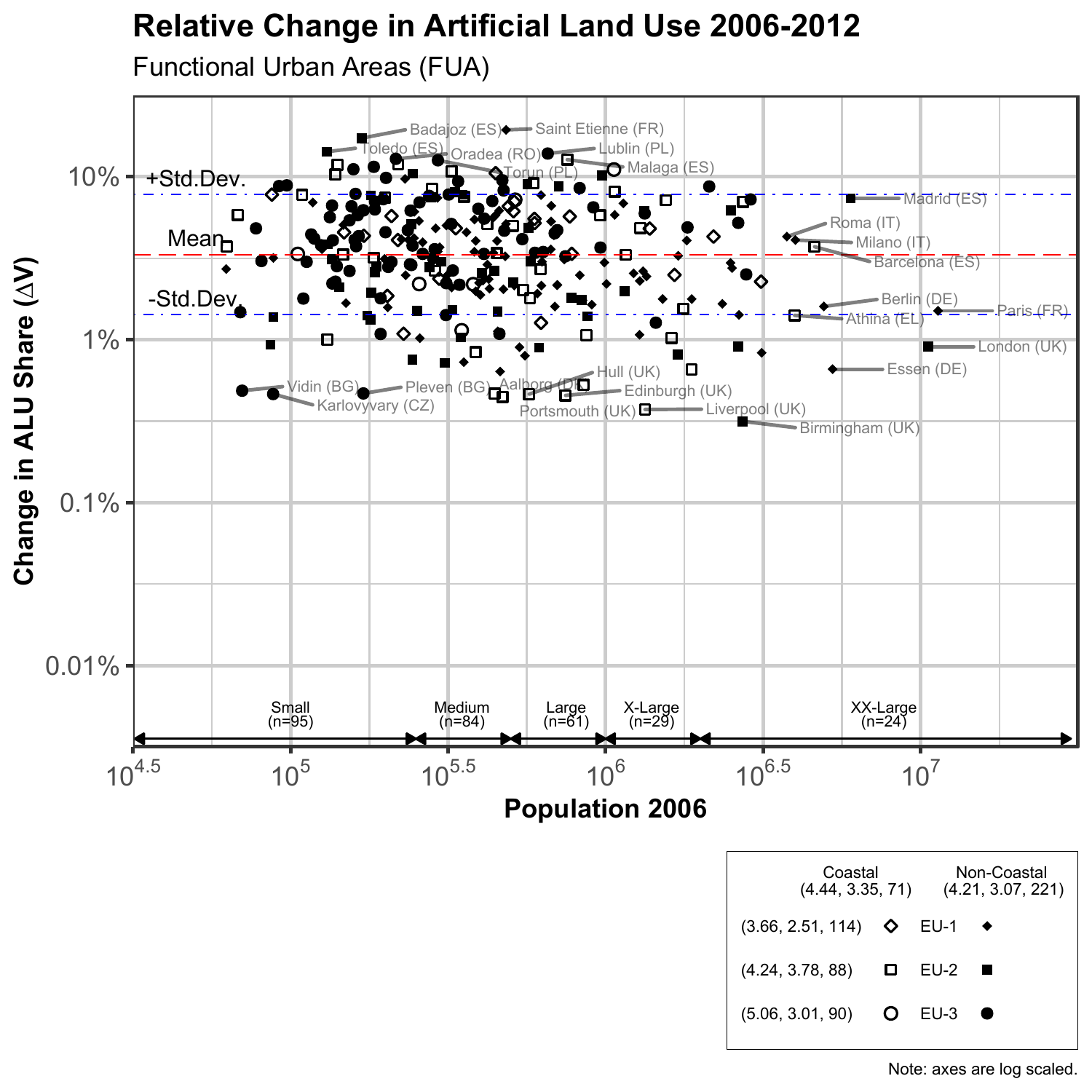}
    \caption{Relative Change in ALU ($\Delta$V) - FUA}
    \label{fig:aggchg}
\end{figure}

Figure \ref{fig:aggchg} shows the relationship between population and ALU change ($\Delta$V) at the FUA level. We can see ALU change ($\Delta$V) is constant across all populations. Most observations are within + or – one standard deviation of the mean, highlighting a narrow range. The largest cities tend to be at the lower end however still within one standard deviation of the mean. The largest cities Paris and London are both below the mean around minus one standard deviation. The third largest city, Madrid, is behaving rather differently to Paris and London but still within the range. Interestingly there appears to be a cluster of former industrial powerhouse cities of Essen, Birmingham and Liverpool to the bottom right. This below average level of artificial land use growth may be a consequence of the decline in certain industry and manufacturing sectors in Europe such as textiles (50\% decrease in production 1995-2015 (Eurostat, 2019)). 

\begin{figure}[H]
    \includegraphics[width=\linewidth]{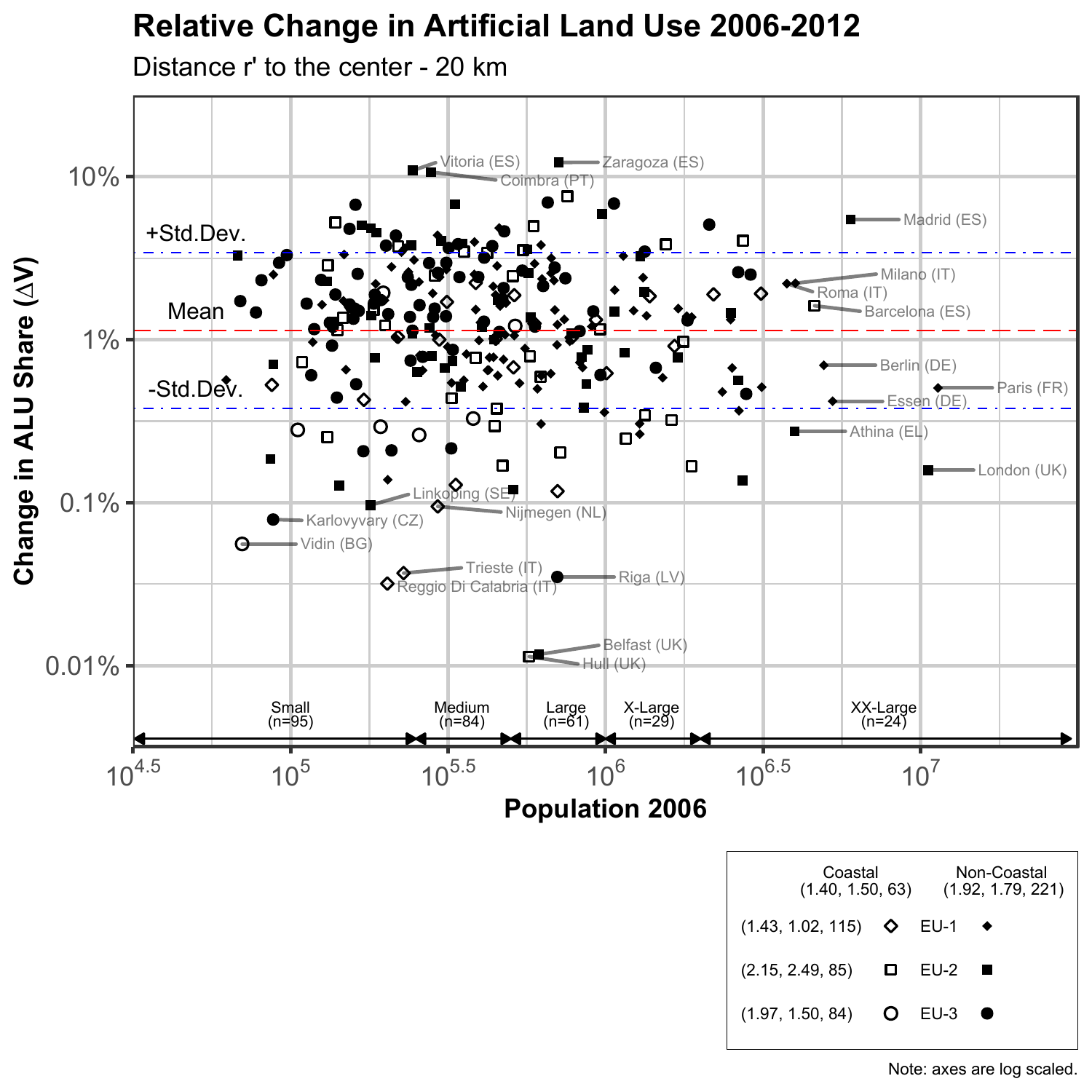}
    \caption{Relative Change in ALU ($\Delta$V) - r$'$=20km}
    \label{fig:rel20}
\end{figure}

\begin{figure}[H]
    \includegraphics[width=\linewidth]{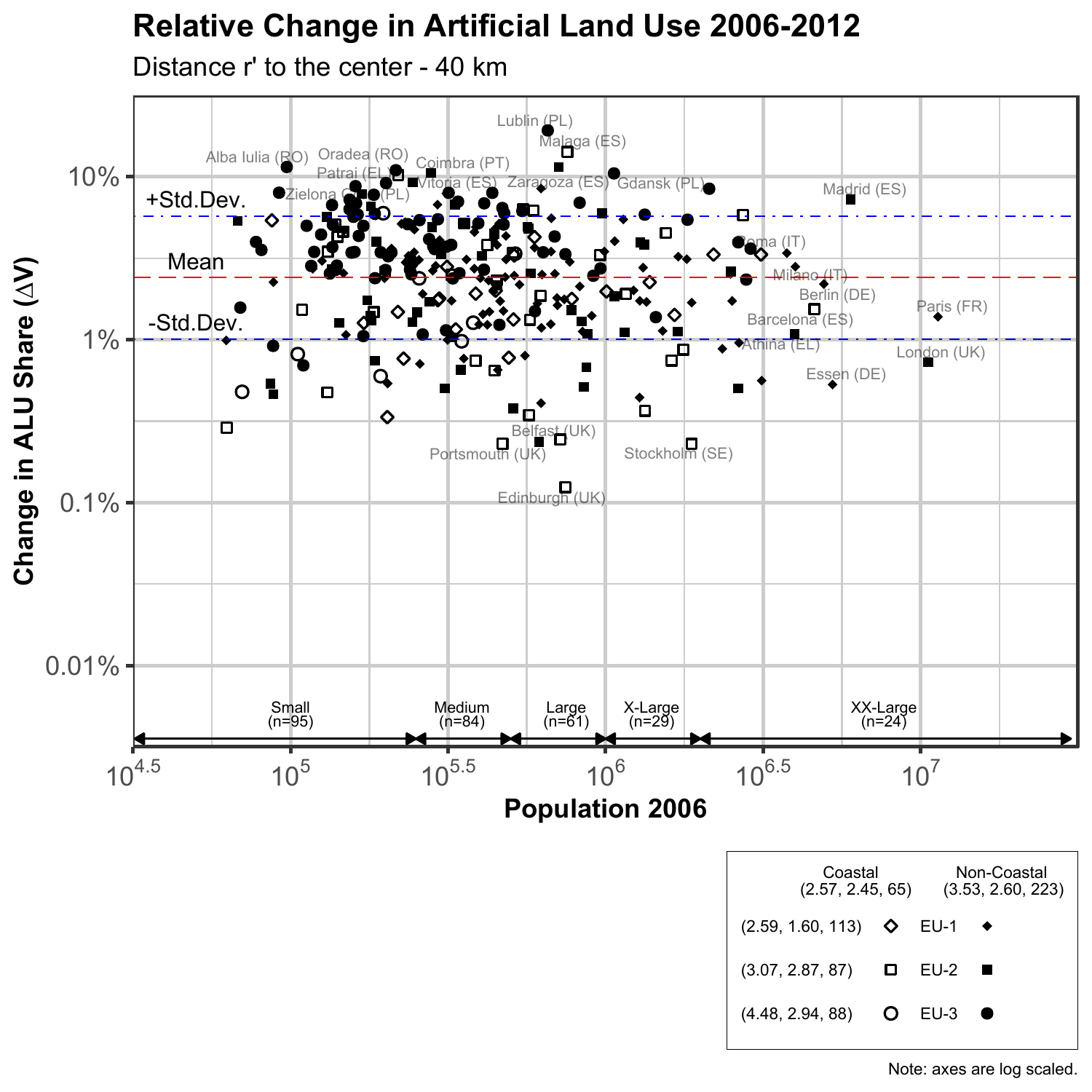}
    \caption{Relative Change in ALU ($\Delta$V) - r$'$=40km}
    \label{fig:rel40}
\end{figure}

Utilising radial discs, the change in ALU ($\Delta$V) is calculated at different distances to the CBD; r$'$=20km and r$'$=40km. In figures \ref{fig:rel20} and \ref{fig:rel40} the log of the change in artificial land use between 2006 and 2012 is plotted against the log of population for 2006 to test whether the change in artificial land use satisfies Gibrat’s Law at different distances to the CBD. In figure \ref{fig:rel20} we examine the internal relative change in ALU share ($\Delta$V) for r$'$=20km. This figure measures total artificial land use within a distance of r$'$=20km to the CBD for 2006 and 2012. The relative change is then computed for each city. Compared to figure \ref{fig:aggchg}, we can see that the mean change is lower ($\approx$1\% compared to $\approx$5\% in the aggregate measure). The change in ALU for cities is also more dispersed than before. There are more cities outside one standard deviation of the mean. This is reflected in the distance between the two standard deviations being larger. Paris has a lower level of change at r$'$=20km however is still within one standard deviation. The change for London has decreased even lower. The low levels for these cities can be explained by them already having a high level of ALU at these distances. There is limited land availability as most available land plots are already urbanisation. The levels of relative change, decreased as a result of examining an area smaller than the FUA.

Comparing figure \ref{fig:rel20} to figure \ref{fig:rel40} we can see the cities are more tightly clustered. This suggests that at r$'$=40km cities are experiencing similar levels of ALU growth/change. There is a group of cities, Cardiff, Belfast and Edinburgh outside one standard deviation of the mean. Given these are all UK cities there may be some economic reasoning behind this. The mean change is also higher and the distance between standard deviations smaller reflecting the fact the cities are more closely clustered. In the r$'$=20km or r$'$=40km graphs the EU group or whether a city is coastal does not appear to have a difference. While the mean value of EU-3 is higher the standard deviation is also higher suggesting there is high within group variability. There are similar results for non-coastal cities. Cities such as Lublin, Madrid and Malaga are constantly around 10\% relative change ($\Delta$V) in all three graphs highlighting a large change has occurred here.

The debate around Gibrat is that it is based upon how we define cities, we show that the rescaling method is robust to the way in which we define cities. At a narrow definition of a city r$'$=20km or at a wider definition r$'$=40km, ALU growth/change ($\Delta$V) and population appear to meet Gibrat’s law. The results highlights the strength of Gibrat’s law for ALU growth ($\Delta$V), as it is robust irrespective of city definition (r$'$=20km, r$'$=40km or FUA). Each of the three graphs yield a similar pattern. Even the use of the FUA definition of cities satisfies Gibrat’s law. City definition based on radial distance to the CBD has highlighted a number of cities at the upper and lower end which were not highlighted when using the FUA definition. These cities warrant further investigation to understand the low change in ALU.

Although, figures \ref{fig:rel20} and \ref{fig:rel40} are characteristic of Gibrat’s law, r$'$=40km corresponds to the law more strongly. As there is a greater variation in the change for r$'$=20km the pattern is more dispersed. Compared to the aggregate measure of ALU change for cities, examining the internal structure of cities has shown that the change is not homogeneous across all distances to the CBD. Relative change in ALU is higher at distances closer to the CBD. This can be explained by the total area of the disc, as distance increases the area required for a 1\% relative change in ALU also increases. Overall the relative change is constant. Even with a more restricted definition of cities, there is some changing of the outliers however Gibrat’s law still holds.

\subsection{Internal structure of cities}

\begin{figure}[H]
    \includegraphics[width=\linewidth]{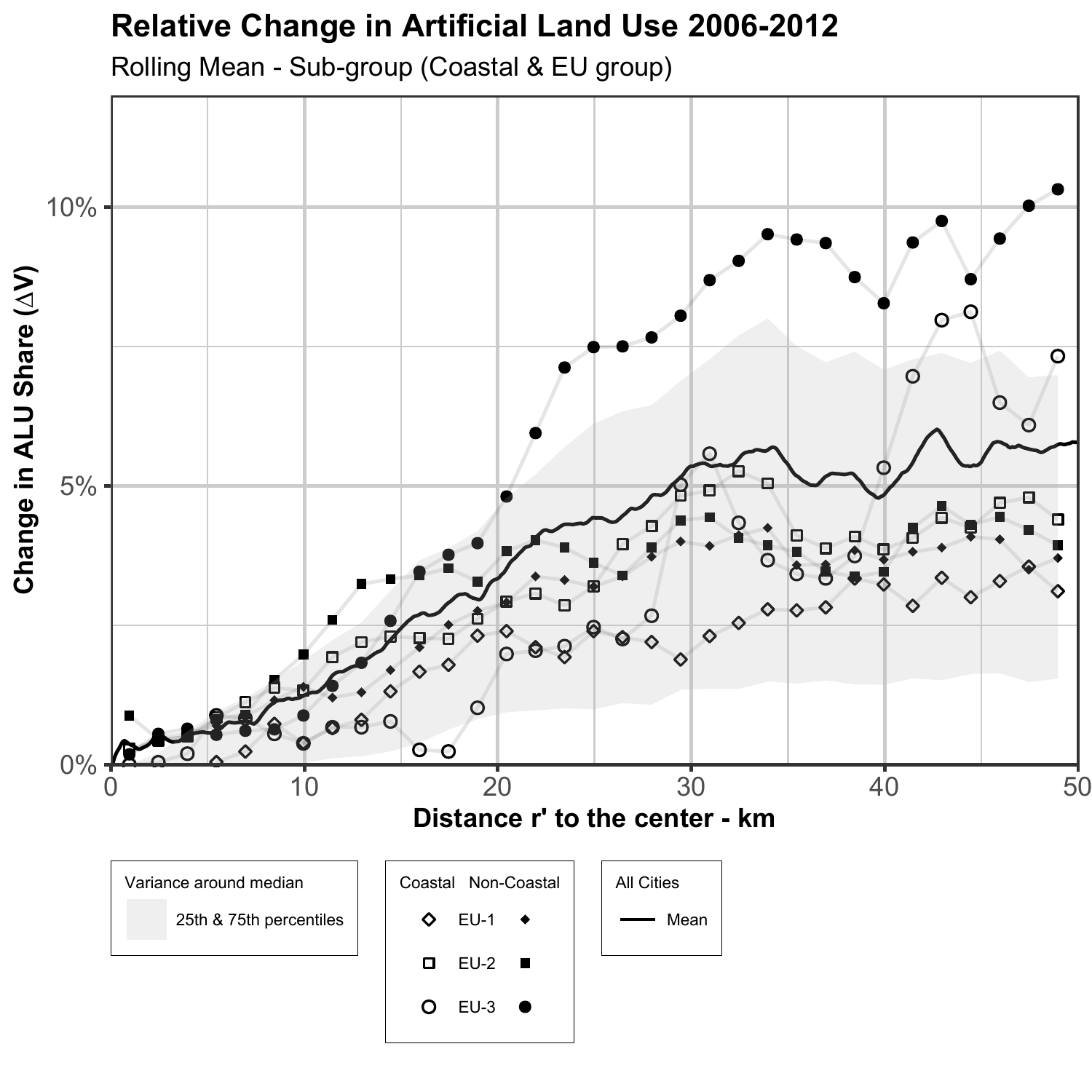}
    \caption{Relative change in ALU (2006-2012) by EU group and coastal status}
    \label{fig:eucst}
\end{figure}

Figure \ref{fig:eucst} shows the relative change in ALU between 2006 and 2012. A rolling mean with a r$'$=2km window is used to reduce the noise in the graph. This is due to the high level of variability in ALU change across rings within a city. Across all rescaled distances, the relative change in ALU is increasing. Mean ALU change is increasing at an increasing rate until a distance of r$'$=30km after which the increase is constantly ~6\%. The graph highlights that ALU change is increasing more in the outer suburbs and periphery. There is more variation at distance rescaled r$'$=10km compared to r$'$=60km. Cities at r$'$=40km are changing at different levels. Some cities are exhibiting greater levels of urban sprawl compared to others. There appears to be a turning point at r$'$=20km beyond which city groups begin to diverge. The non-coastal EU-3 have the highest levels of ALU change beyond r$'$=20km with a growth rate greater than the \nth{75} percentile on average. The ALU change for EU-3 coastal cities varies, this can be explained by the low number of cities (n=6) in this category.  The lowest levels of growth was experienced in EU-1 coastal cities, the difference between EU-1 coastal and non-coastal highlights the impact of topography on ALU change. For all city categories a change in the rate of growth occurs at r$'$=20km.

\begin{figure}[H]
    \includegraphics[width=\linewidth]{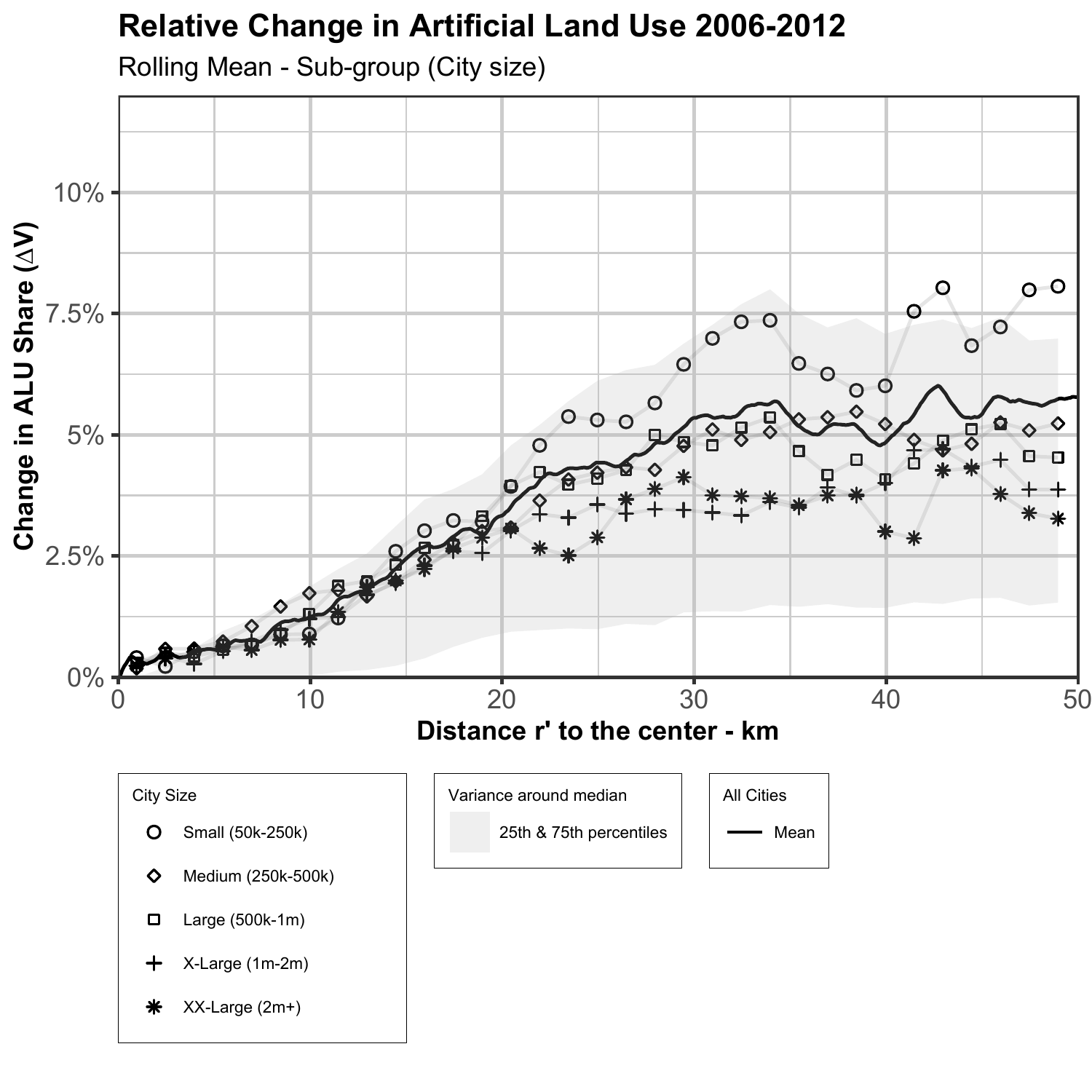}
    \caption{Relative change in ALU (2006-2012) grouped by city size}
    \label{fig:size}
\end{figure}

In figure \ref{fig:size} cities are grouped by city size categories. At distances r$'$=0-20km cities on average have similar levels of ALU change. Beyond a distance of r$'$=20km there is a divergence in the rate of change for cities with small/medium cities growing on average more compared to xx-large cities. Focusing on r$'$=40km we see a clear sorting of the city size categories from smallest to largest. There is a slight deviation from Gibrat’s law with small/medium cities growing on average more compared to larger cities. To examine this finding further the conversion rate of non-urbanised land to ALU is examined. Larger cities may already have higher levels of ALU compared to smaller cities which would reduce the quantity of land available for development.

\begin{figure}[H]
    \includegraphics[width=\linewidth]{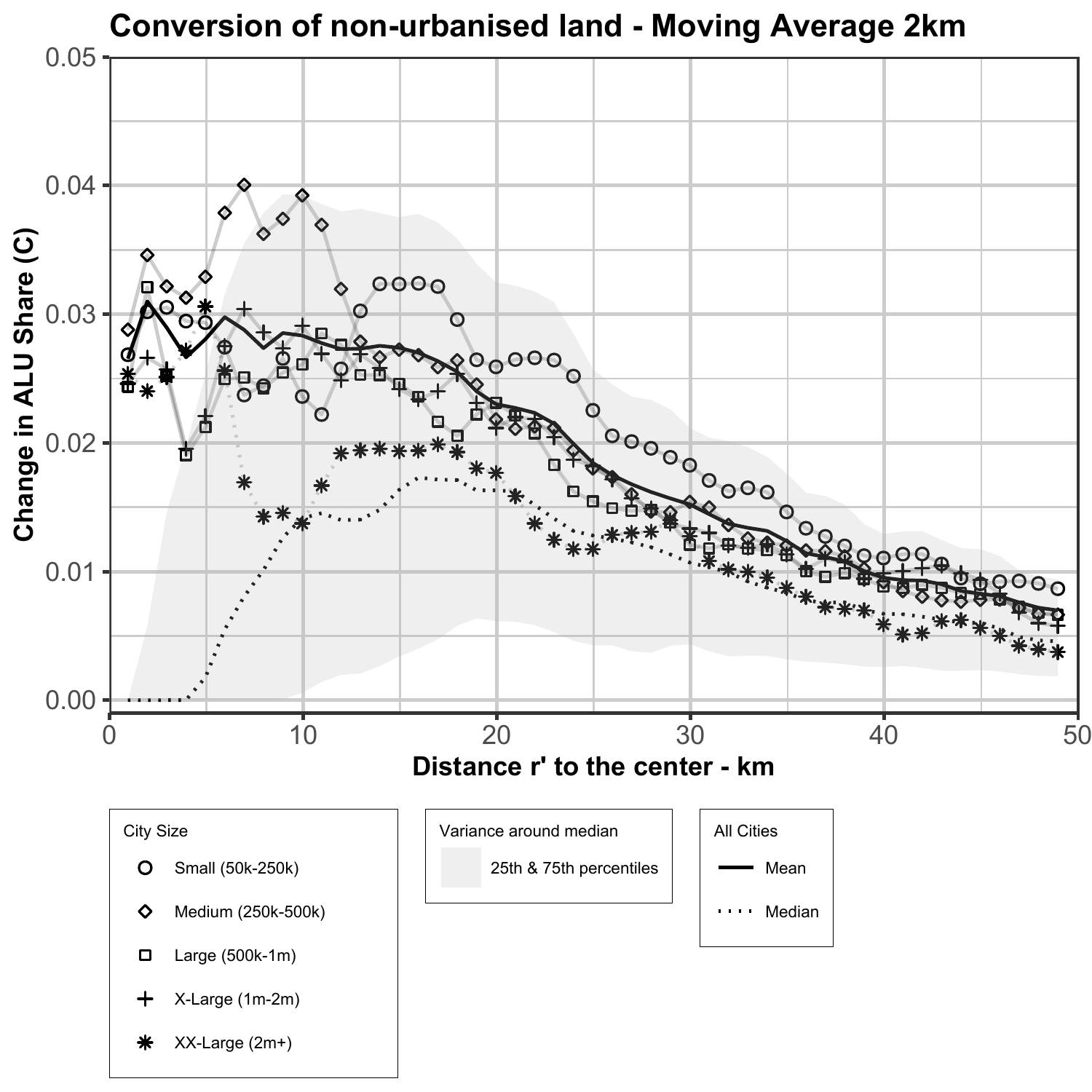}
    \caption{Conversion of non-urbanised land to ALU}
    \label{fig:conv}
\end{figure}

Figure \ref{fig:conv} shows the moving average of the conversion rate of non-urbanised land. Although the relative change in ALU is increasing for all distances, the increase as a share of non-urbanised land decreases with distance to the CBD. At distances less than r$'$=12km there is high variation between the city size categories, beyond this distance however again there is a sorting of the city size categories. Beyond r$'$=12km small/medium cities are on average experiencing greater levels of urban expansion. This suggests that these smaller cities are sprawling more relative to xx-large cities such as Paris or London. At r$'$ $<$5km there is little to no conversion of non-urbanised land for the vast majority of cities. Although redevelopment of existing urbanised land may occur, the potential to develop non-urbanised land is limited. 

\section{Conclusion}
Our results have shown that a kind of Gibrat’s law holds for urban land use change. However, there is a small deviation as larger cities grow slightly less compared to smaller ones. The aggregate measures of ALU growth/change ($\Delta$V) are shown to follow Gibrat and are robust to changes in the way we define a city. Results are similar whether the aggregate FUA, r$'$=20km (reflecting the city core) or r$'$=40km (considering the suburbs) is used. Using a tight definition of a city as in the r$'$=20km case, we see greater variability in the level of ALU change. This may reflect the differences across cities in demand for land in the city core.

There is however a need to focus on the internal structure and internal change of ALU across cities. Utilising the scaling law found by \citet{Lemoy2018} has enabled us to examine the internal structure of cities despite cities having various populations and extents. Cities are more or less growing in a consistent way in the core however in the periphery there are clearly different growth rates. Grouping cities based on city size and examining the intra city growth rate, there are clear differences between city size groups with smaller cities having a higher ALU growth rate. These differences however only become apparent beyond r$'$=20km. New EU member states (EU-3) grew at a faster rate compared to older more established member states (EU-1 and EU-2). The reasons behind these differences warrant further investigation.

The overall trend of the internal structure of cities shows increasing levels of ALU at all distances beyond r$'$=20km. These increasing levels of urban expansion point to increasing levels of urban sprawl with more sprawl occurring in smaller cities. This raises important questions around the sustainability of cities as this evidence points to increasing urban sprawl and stagnant growth in urban centres across cities of all sizes. It also bears theoretical implications on the nature of sprawl and its scaling.

Turning to the conversion of non-urbanised land, there is a differentiation between large and small cities around the intensity of the change and where this change occurs. Large cities appear to have a greater level of conversion at the core compared to smaller cities, beyond the core smaller cities have greater levels of conversion. Due to the demand for land in larger cities, there would appear to be a greater incentive to utilise any non-urbanised land. For smaller cities this is perhaps not as large an issue as land in the periphery can be easily sourced and the actual distance to the CBD remains relatively small.

The results highlight issues in relation to how we define a city and more specifically its outer limit. Using a narrow versus a wider definition, we are able to see the level of variability in ALU change. A recursive definition for cities is required to remove away from subjectivity defining city boundaries/limits. Changes in fractal dimension of built up areas may also offer further insights into the boundary used for a city \citep{frankhauser1998fractal,tannier2011fractal,tannier2013defining}. Scaling has the potential to be the solution in this regard, enabling us to determine a city’s extent through the use of data.

\bibliography{bookch}

\end{document}